# Lindblad master equation for the damped harmonic oscillator with deformed dissipation

A. Isar†‡[(a)], A. Sandulescu† and W. Scheid‡

†Department of Theoretical Physics, Institute of Physics and Nuclear Engineering

Bucharest-Magurele, Romania

‡Institut für Theoretische Physik der Justus-Liebig-Universität

Giessen, Germany

**Abstract**

In the framework of the Lindblad theory for open quantum systems, a master equation for the quantum harmonic oscillator interacting with a dissipative environment, in particular with a thermal bath, is derived for the case when the interaction is based on deformed algebra. The equations of motion for observables strongly depend on the deformation function. The expectation values of the number operator and squared number operator are calculated in the limit of a small deformation parameter for the case of zero temperature of the thermal bath. The steady state solution of the equation for the density matrix in the number representation is obtained and its independence of the deformation is shown.

## 1 Introduction

For more than a decade an interest has been consisting for the study of deformations of Lie algebras – so-called quantum algebras or quantum groups, whose structure produced important results and consequences in several branches of physics [1, 2]. Their use in physics became intense with the introduction in 1989, by Biedenharn [3] and MacFarlane [4], of the $q$-deformed Heisenberg-Weyl algebra ($q$-deformed quantum harmonic oscillator). There are, at least, two properties which make $q$-deformed oscillators interesting objects for physics. The first is the fact that they naturally appear as the basic building blocks of completely integrable theories [5]. The second concerns the connection between $q$-deformation and nonlinearity. In Refs. [6, 7, 8] there have been considered some aspects of the physical nature of the $q$-oscillator and it was shown that it is subject to nonlinear vibrations with a dependence of the frequency on the amplitude. The $q$-deformed Bose distribution was obtained and shown how the $q$-nonlinearity produces a changed Planck distribution formula [6, 7, 9].

In the present paper we study the connection of quantum deformation and quantum dissipation, by setting a model for a harmonic oscillator interacting with a dissipative environment, where the interaction between both is $f$- and $q$-deformed. The master equation which describes the dynamics of the oscillator represents a $f$-, respectively $q$-deformed version of the master equation obtained in the framework of the Lindblad theory for open quantum systems [10]. When the deformation vanishes, we recover the Lindblad master equation for the damped harmonic oscillator [11, 12]. For a certain choice of the environment coefficients, a master equation for the damped deformed oscillator has also been derived by Mancini [13]. In Ref. [14], Ellinas used the $q$-deformed oscillator for treatments of dissipation of a two-level atom and of a laser mode. In a previous paper [15], using a variant of the Mancini's model [13], we derived a master equation for the $f$-deformed oscillator in the presence of a dissipative environment.

The paper is organized as follows. In Sec. 2 we remind the basics about the $f$-deformed quantum oscillator, in particular the $q$-oscillator. In Sec. 3, using the Lindblad theory for open quantum systems, we derive a master equation for the harmonic oscillator in the presence of a dissipative environment, for the case when the interaction is deformed. The equations of motion obtained for different observables present a strong dependence on the deformation. In Sec. 4 we solve a linear system of two coupled first order differential equations for the expectation values of the number operator and squared number operator, in the limit of small deformation parameter and for zero temperature of the thermal bath. Then in Sec. 5 we write in the number representation the equation for the density matrix and find the stationary state. In the particular case when the environment is a thermal bath we obtain the steady-state Boltzmann distribution of the ordinary harmonic oscillator, in which the deformation does not play any role. A summary is given in Sec. 6.

## 2  $f$- and $q$-deformed quantum oscillators

It is known that the ordinary operators $\{1, a, a^\dagger, N\}$ form the Lie algebra of the Heisenberg-Weyl group and the linear harmonic oscillator can be connected with the generators of the Heisenberg-Weyl Lie group. The $f$-deformed quantum oscillators [16] are defined as the algebra generated by the operators $\{1, A, A^\dagger, N\}$. The $f$-deformed oscillator operators are given as follows [16]:

$$A = af(N) = f(N+1)a, \quad A^\dagger = f(N)a^\dagger = a^\dagger f(N+1), \tag{1}$$

where $N = a^\dagger a$. They satisfy the commutation relations

$$[A, N] = A, \quad [A^\dagger, N] = -A^\dagger \tag{2}$$

and

$$[A, A^\dagger] = (N+1)f^2(N+1) - Nf^2(N). \tag{3}$$

The function $f$, which is characteristics for the deformation, has a dependence on the deformation parameter such that when the deformation disappears, then $f \to 1$ and the usual algebra is recovered. Without loss of generality, $f$ can be chosen real and nonnegative and it is reasonable from the physical point of view to assume [8] that $f(0) = 1$ and $f(N) = 1$ for a suitable large $N$. The transformation (1) from the operators $a, a^\dagger$ to $A, A^\dagger$ represents a nonlinear non-canonical transformation, since it does not preserve the commutation relation. On the other side, the transformation (1) does not change the dynamics, i. e. the equations of motion and, therefore, the transformation (1) contains a symmetry for these equations.

The notion of $f$-oscillators generalizes the notion of $q$-oscillators. Indeed, for

$$f(N) = \sqrt{\frac{[N]}{N}}, \quad [N] \equiv \frac{q^N - q^{-N}}{q - q^{-1}}, \tag{4}$$

where $q$ is the deformation parameter (a dimensionless $c$-number), the operators $A$ and $A^\dagger$ in Eqs. (1) become the $q$-deformed boson annihilation and creation operators [3, 4]. For the $q$-deformed harmonic oscillator the commutation relation (3) becomes

$$[A, A^\dagger] = [N+1] - [N]. \tag{5}$$

If $q$ is real, then

$$[N] = \frac{\sinh(N \ln q)}{\sinh(\ln q)} \tag{6}$$

and the condition of Hermitian conjugation $(A^\dagger)^\dagger = A$ is satisfied. In the limit $q \to 1$, $q$-deformed operators tend to the ordinary operators, $\lim_{q \to 1}[N] = N$ and Eq. (5) goes to the usual boson commutation relation $[A, A^\dagger] = 1$. The $q$-deformed boson operators $A$ and $A^\dagger$ can be expressed in terms of the usual boson operators $a$ and $a^\dagger$ (satisfying $[a, a^\dagger] = 1$, $N = a^\dagger a$ and $[a, N] = a$, $[a^\dagger, N] = -a^\dagger$) through the relations [17, 18]:

$$A = \sqrt{\frac{[N+1]}{N+1}} a = a \sqrt{\frac{[N]}{N}}, \quad A^\dagger = a^\dagger \sqrt{\frac{[N+1]}{N+1}} = \sqrt{\frac{[N]}{N}} a^\dagger. \tag{7}$$

Using a nonlinear map [19, 20], the $q$-oscillator has been interpreted [6, 7] as a nonlinear oscillator with a special type of nonlinearity which classically corresponds to an energy dependence of the oscillator frequency. Other nonlinearities can also be introduced by making the frequency to depend, through the deformation function $f$, on other constants of motion, different from energy [6, 16]. Other examples of deformation functions $f$ can be found in Refs. [21, 22, 23, 24].

# 3 Lindblad master equation with deformed dissipation

Dissipative processes imply irreversibility, which means a preferred direction in time. Therefore, it is generally thought that a rigorous introduction of dissipation into a quantum system can be done in the framework of quantum dynamical semigroups [10, 25, 26]. The most general form of the generators of such semigroups was given by Lindblad [10]. This formalism has thoroughly been studied for the damped harmonic oscillator [11, 12, 27, 28] and applied to various physical phenomena, for instance, damping of collective modes in deep inelastic collisions in nuclear physics [12, 29].

According to the axiomatic theory of Lindblad [10], the most general quantum Markovian master equation for the density operator $\rho$ has the following form:

$$\frac{d\rho(t)}{dt} = -\frac{i}{\hbar}[H, \rho(t)] + \frac{1}{2\hbar}\sum_j ([V_j\rho(t), V_j^\dagger] + [V_j, \rho(t)V_j^\dagger]), \qquad (8)$$

where $H$ is the Hamiltonian of the system and the operators $V_j$ and $V_j^\dagger$, defined on the Hilbert space of $H$ describe the interaction between system and environment.

In this paper we consider that the open system is a harmonic oscillator with the Hamiltonian

$$H = \frac{1}{2m}p^2 + \frac{m\omega^2}{2}q^2 \qquad (9)$$

and the operators $V_j$ are taken first degree polynomials in the basic observables $p$ and $q$ of the one-dimensional quantum system. Because in the linear space of these polynomials the operators $p$ and $q$ give a basis, there exist only two **C**-linear independent operators $V_1, V_2$ which have the form

$$V_j = a_j p + b_j q, \ j = 1, 2, \qquad (10)$$

with $a_j, b_j$ complex numbers [27]. With these choices and introducing the annihilation and creation operators via the relations

$$q = \sqrt{\frac{\hbar}{2m\omega}}(a^\dagger + a), \ \ p = i\sqrt{\frac{\hbar m\omega}{2}}(a^\dagger - a), \qquad (11)$$

the master equation (8) takes the form

$$\frac{d\rho}{dt} = -\frac{i}{\hbar}[H, \rho] + \frac{1}{2}\{D_1(a^\dagger a^\dagger \rho + \rho a^\dagger a^\dagger - 2a^\dagger \rho a^\dagger)$$
$$+(D_2 + \lambda)(a\rho a^\dagger - a^\dagger a \rho) + (D_2 - \lambda)(a^\dagger \rho a - \rho a a^\dagger) + \text{H.c.}\}, \qquad (12)$$

where $H = (1/2)\hbar\omega(aa^\dagger + a^\dagger a)$ and

$$D_1 \equiv \frac{1}{\hbar}(m\omega D_{qq} - \frac{D_{pp}}{m\omega} + 2iD_{pq}), \quad D_2 \equiv \frac{1}{\hbar}(m\omega D_{qq} + \frac{D_{pp}}{m\omega}). \quad (13)$$

We used the notations:

$$D_{qq} = \frac{\hbar}{2}\sum_{j=1,2}|a_j|^2, \quad D_{pp} = \frac{\hbar}{2}\sum_{j=1,2}|b_j|^2, \quad D_{pq} = D_{qp} = -\frac{\hbar}{2}\text{Re}\sum_{j=1,2}a_j^*b_j, \quad \lambda = -\text{Im}\sum_{j=1,2}a_j^*b_j, \quad (14)$$

where $D_{pp}, D_{qq}$ and $D_{pq}$ are the diffusion coefficients and $\lambda$ the dissipation constant. These quantities satisfy the following fundamental constraints [11, 12]:

$$\text{i) } D_{pp} > 0, \quad \text{ii) } D_{qq} > 0, \quad \text{iii) } D_{pp}D_{qq} - D_{pq}^2 \geq \frac{\hbar^2\lambda^2}{4}. \quad (15)$$

In the particular case when the asymptotic state is a Gibbs state

$$\rho_G(\infty) = e^{-\frac{H}{kT}}/\text{Tr}e^{-\frac{H}{kT}}, \quad (16)$$

the diffusion coefficients reduce to

$$D_{pp} = \frac{\lambda}{2}\hbar m\omega\coth\frac{\hbar\omega}{2kT}, \quad D_{qq} = \frac{\lambda}{2}\frac{\hbar}{m\omega}\coth\frac{\hbar\omega}{2kT}, \quad D_{pq} = 0, \quad (17)$$

where $T$ is the temperature of the thermal bath and $k$ the Boltzmann constant. The relations (15) ensure the positivity of the density operator. Due to the dissipative terms in Eq. (12), the energy of the open system is dissipated into the environment, but noise arises also (in particular, thermal noise), since the environment also distributes some of its energy back to the system.

Next we propose a model where we deform the dissipative part in Eq. (12) by replacing the operators $a$ and $a^\dagger$ by the $f$-deformed operators (1). Then the master equation (12) becomes:

$$\frac{d\rho}{dt} = -i\omega[N, \rho]$$
$$+\frac{1}{2}D_1\{f(N-1)f(N)a^\dagger a^\dagger\rho + \rho a^\dagger a^\dagger f(N+1)f(N+2) - 2f(N)a^\dagger\rho a^\dagger f(N+1)\}$$
$$+\frac{1}{2}D_1^*\{f(N+1)f(N+2)aa\rho + \rho aaf(N-1)f(N) - 2f(N+1)a\rho af(N)\}$$
$$-\frac{1}{2}(D_2 + \lambda)\{Nf^2(N)\rho + \rho Nf^2(N) - 2f(N+1)a\rho a^\dagger f(N+1)\}$$
$$-\frac{1}{2}(D_2 - \lambda)\{(N+1)f^2(N+1)\rho + \rho(N+1)f^2(N+1) - 2f(N)a^\dagger\rho af(N)\}. \quad (18)$$

We stress that in the proposed model we do not introduce the deformation into the commutator containing the oscillator Hamiltonian $H$, but we do it only in the dissipative part of the master equation, which describes the influence of the environment on the oscillator. The deformation of the oscillator Hamiltonian would imply the deformation of the diffusion coefficients too and this model has been treated in our previous paper [15]. Obviously, the obtained master equation preserves the Hermiticity property of the density operator and the normalization (unit trace) at all times (if at the initial time it has these properties). In the limit $f \to 1$ the deformation disappears and Eq. (18) becomes the Markovian master equation for the usual damped harmonic oscillator, obtained in the Lindblad theory for open quantum systems. In the particular case of a thermal equilibrium of the bath at temperature $T$ with diffusion coefficients (17), the master equation (18) takes the form

$$\frac{d\rho}{dt} = -i\omega[N,\rho]$$
$$-\frac{1}{2}\lambda(\coth\frac{\hbar\omega}{2kT}+1)\{Nf^2(N)\rho + \rho N f^2(N) - 2f(N+1)a\rho a^\dagger f(N+1)\}$$
$$-\frac{1}{2}\lambda(\coth\frac{\hbar\omega}{2kT}-1)\{(N+1)f^2(N+1)\rho + \rho(N+1)f^2(N+1) - 2f(N)a^\dagger\rho a f(N)\}. \quad (19)$$

If the bath temperature is $T=0$, the master equation (19) simplifies and becomes

$$\frac{d\rho}{dt} = -i\omega[N,\rho] - \lambda\{Nf^2(N)\rho + \rho N f^2(N) - 2f(N+1)a\rho a^\dagger f(N+1)\}. \quad (20)$$

## 4 Equations of motion for $<N>$ and $<N^2>$

The meaning of the master equation becomes clear when we transform it into equations satisfied by various expectation values of observables involved in the master equation, $<B> = \text{Tr}[\rho(t)B]$, where $B$ is the operator corresponding to such an observable. We give an example, multiplying both sides of Eq. (19) by the number operator $N$ and taking the trace. Then we obtain the following equation of motion for the expectation value of $N$:

$$\frac{d}{dt}<N> = \lambda\{(\coth\frac{\hbar\omega}{2kT}-1)<(N+1)f^2(N+1)> - (\coth\frac{\hbar\omega}{2kT}+1)<Nf^2(N)>\}. \quad (21)$$

We consider another example, namely for the operator $N^2$ and we obtain the following equation:

$$\frac{d}{dt}<N^2> = \lambda\{(\coth\frac{\hbar\omega}{2kT}-1)<(2N+1)(N+1)f^2(N+1)>$$
$$-(\coth\frac{\hbar\omega}{2kT}+1)<(2N-1)Nf^2(N)>\}. \quad (22)$$

We observe that the equations of motion for the expectation values contain complicated nonlinearities introduced by the deformation function $f$ and, in addition, they do not form a closed system. In order to introduce some procedure for truncating the set of these coupled equations, we consider the simplest case of deformation – the $q$-deformation. In this case, using Eq. (4), Eqs. (21) and (22) take the following form:

$$\frac{d}{dt}<N> = \lambda\{(\coth\frac{\hbar\omega}{2kT} - 1)<[N+1]> - (\coth\frac{\hbar\omega}{2kT} + 1)<[N]>\}, \qquad (23)$$

$$\frac{d}{dt}<N^2> = \lambda\{(\coth\frac{\hbar\omega}{2kT} - 1)<(2N+1)[N+1]>$$
$$-(\coth\frac{\hbar\omega}{2kT} + 1)<(2N-1)[N]>\}. \qquad (24)$$

Now we write the $q$-operator $[N]$ (6) in the lowest order of the Taylor expansion in the small deformation parameter $\tau$ (we consider the case of real $q = \exp\tau$) [2]:

$$[N] = N - \frac{\tau^2}{6}(N - N^3). \qquad (25)$$

With this expression, Eqs. (23) and (24) become:

$$\frac{d}{dt}<N> = \lambda\{\coth\frac{\hbar\omega}{2kT} - 1 - 2<N>$$
$$+\frac{\tau^2}{6}[3\coth\frac{\hbar\omega}{2kT}<N(N+1)> - <2N^3 + 3N^2 + N>]\}, \qquad (26)$$

$$\frac{d}{dt}<N^2> = \lambda\{\coth\frac{\hbar\omega}{2kT}<4N+1> - <4N^2 + 2N + 1>$$
$$+\frac{\tau^2}{6}[\coth\frac{\hbar\omega}{2kT}<8N^3 + 9N^2 + N> - <4N^4 + 6N^3 + 5N^2 + 3N>]\}. \qquad (27)$$

Now we make use of the relation

$$N^3 = a^{\dagger 3}a^3 + 3N^2 - 2N \qquad (28)$$

and make the assumption that $<a^{\dagger 3}a^3>$ can be neglected for not too high values of the number operator $N$. Then, replacing $N^3$ by $3N^2 - 2N$ in Eqs. (26) and (27), we obtain:

$$\frac{d}{dt}<N> = \lambda\{\frac{\tau^2}{2}(\coth\frac{\hbar\omega}{2kT} - 3)<N^2>$$
$$-[2 - \frac{\tau^2}{2}(\coth\frac{\hbar\omega}{2kT} + 1)]<N> + \coth\frac{\hbar\omega}{2kT} - 1\}, \qquad (29)$$

$$\frac{d}{dt}<N^2> = \lambda\{[\frac{\tau^2}{2}(11\coth\frac{\hbar\omega}{2kT} - 17) - 4]<N^2>$$
$$+[4\coth\frac{\hbar\omega}{2kT} - 2 - \frac{\tau^2}{2}(5\coth\frac{\hbar\omega}{2kT} - 11)]<N> + \coth\frac{\hbar\omega}{2kT} - 1\}. \qquad (30)$$

By this truncation, we obtained a closed non-homogeneous linear system of first order coupled differential equations for the expectation values $<N>$ and $<N^2>$, which can analytically be solved. For simplicity, we will solve this system in the case of the temperature $T = 0$ of the thermal bath. Then Eqs. (29) and (30) become:

$$\frac{d}{dt} <N> = -\tau^2 \lambda <N^2> + (\tau^2 - 2)\lambda <N>, \tag{31}$$

$$\frac{d}{dt} <N^2> = -(3\tau^2 + 4)\lambda <N^2> + (2 + 3\tau^2)\lambda <N>. \tag{32}$$

The integration of this system is straightforward. Introducing the vector

$$S(t) = \begin{pmatrix} <N(t)> \\ <N^2(t)> \end{pmatrix} \tag{33}$$

and the matrix

$$M = \lambda \begin{pmatrix} \tau^2 - 2 & -\tau^2 \\ 2 + 3\tau^2 & -3\tau^2 - 4 \end{pmatrix}, \tag{34}$$

the system (31) – (32) can be written

$$\frac{dS(t)}{dt} = MS(t). \tag{35}$$

Now $M$ can be written as $M = R^{-1} F R$ with $F$ the diagonal matrix that has as elements the roots of the characteristic equation of the linear system. It follows that the solution of the system is

$$S(t) = R^{-1} e^{Ft} R S(0). \tag{36}$$

We obtain

$$R^{-1} e^{Ft} R = \frac{1}{2(1+\tau^2)} \begin{pmatrix} (2+3\tau^2)e^{-2\lambda t} - \tau^2 e^{-2\lambda(2+\tau^2)t} & -\tau^2(e^{-2\lambda t} - e^{-2\lambda(2+\tau^2)t}) \\ (2+3\tau^2)(e^{-2\lambda t} - e^{-2\lambda(2+\tau^2)t}) & -\tau^2 e^{-2\lambda t} + (2+3\tau^2)e^{-2\lambda(2+\tau^2)t} \end{pmatrix} \tag{37}$$

and the solutions are the following:

$$<N(t)> = \frac{1}{2(1+\tau^2)} \{-\tau^2(e^{-2\lambda t} - e^{-2\lambda(2+\tau^2)t}) <N^2(0)>$$
$$+ [(2+3\tau^2)e^{-2\lambda t} - \tau^2 e^{-2\lambda(2+\tau^2)t}] <N(0)>\}, \tag{38}$$

$$<N^2(t)> = \frac{1}{2(1+\tau^2)} \{[-\tau^2 e^{-2\lambda t} + (2+3\tau^2)e^{-2\lambda(2+\tau^2)t}] <N^2(0)>$$
$$+ (2+3\tau^2)(e^{-2\lambda t} - e^{-2\lambda(2+\tau^2)t}) <N(0)>\}, \tag{39}$$

or, in a more simple form

$$< N(t) >= \frac{1}{2(1+\tau^2)}\{e^{-2\lambda t}[-\tau^2 < N^2(0) > +(2+3\tau^2) < N(0) >]$$
$$-\tau^2 e^{-2\lambda(2+\tau^2)t}(< N(0) > - < N^2(0) >)\}, \qquad (40)$$

$$< N^2(t) >= \frac{1}{2(1+\tau^2)}\{[e^{-2\lambda t}[-\tau^2 < N^2(0) > +(2+3\tau^2) < N(0) >]$$
$$+(2+3\tau^2)e^{-2\lambda(2+\tau^2)t}(< N^2(0) > - < N(0) >)\}. \qquad (41)$$

Keeping only the relevant terms in small $\tau^2$, we can write the solutions as follows:

$$< N(t) >= e^{-4\lambda t}\frac{\tau^2}{2}(< N^2(0) > - < N(0) >)$$
$$-e^{-2\lambda t}[\frac{\tau^2}{2} < N^2(0) > -(1+\frac{\tau^2}{2}) < N(0) >], \qquad (42)$$

$$< N^2(t) >= e^{-4\lambda t}(1 + \frac{\tau^2}{2} - 2\lambda\tau^2 t)(< N^2(0) > - < N(0) >)$$
$$-e^{-2\lambda t}[\frac{\tau^2}{2} < N^2(0) > -(1+\frac{\tau^2}{2}) < N(0) >]. \qquad (43)$$

In the case when $q$ is a phase ($q = \exp(i\tau), \tau$ real), the expectation values of the number operator and squared number operator are obtained by making the replacement $\tau^2 \to -\tau^2$ in the corresponding expressions (42) and (43). In the limit of long times we obtain $< N^2(\infty) >=< N(\infty) >= 0$, like in the case of the undeformed case. This is as we expected, since we are in the case of $T = 0$. In the limit of vanishing deformation ($\tau \to 0$) we obtain:

$$< N(t) >= e^{-2\lambda t} < N(0) >, \qquad (44)$$

$$< N^2(t) >= e^{-4\lambda t}(< N^2(0) > - < N(0) >) + e^{-2\lambda t} < N(0) >, \qquad (45)$$

in concordance with the previous results obtained in the study of the damped harmonic oscillator in the Lindblad theory [30]. The exponential decreasing of the expectation values of the number operator and of the squared number operator are represented graphically in Fig. 1 for both the cases when $q = \exp\tau$ and $q = \exp(i\tau)$, for a small real deformation parameter $\tau = \sqrt{0.2}$. For comparison we represented also the evolution of these observables in the case when the deformation is absent. The equations of motion for other observables contain also nonlinearities introduced by the deformation function $f$, like in the considered case.

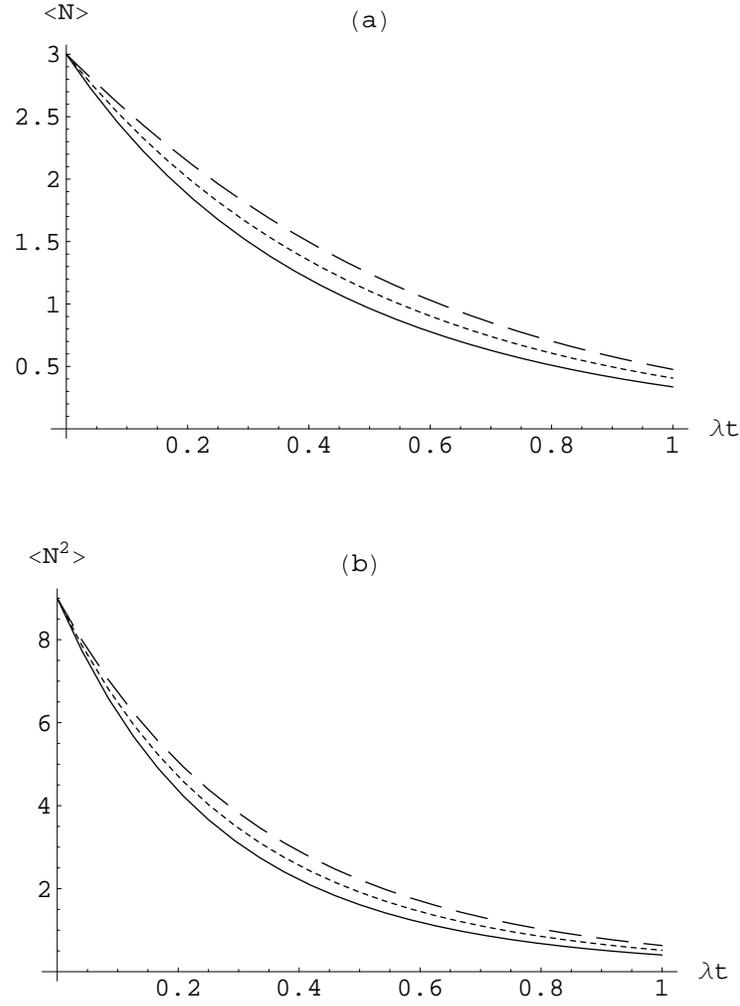

Figure 1: Dependence of (a) averaged number operator $<N>$ given by Eq. (42) and of (b) averaged squared number operator $<N^2>$ given by Eq. (43) on the dimensionless time $\lambda t$, for $<N(0)>=3, <N^2(0)>=9$ and $\tau^2=0.2$. The cases of $q=\exp\tau$, $q=\exp(i\tau)$ with $\tau$ real and $q=0$ are represented by solid, dashed and dotted lines, respectively.

# 5  Equation for the density matrix

Let us rewrite the master equation (18) for the density matrix by means of the number representation. Namely, we take the matrix elements of each term between different number states denoted by $|n>$, and using $N|n> = n|n>$, $a^+|n> = \sqrt{n+1}|n+1>$ and $a|n> = \sqrt{n}|n-1>$, we get

$$\frac{d\rho_{mn}}{dt} = -i\omega(m-n)\rho_{mn}$$
$$-\frac{1}{2}\{(D_2+\lambda)(mf^2(m)+nf^2(n)) + (D_2-\lambda)((m+1)f^2(m+1)+(n+1)f^2(n+1))\}\rho_{mn}$$
$$+(D_2+\lambda)\sqrt{(m+1)(n+1)}f(m+1)f(n+1)\rho_{m+1,n+1}$$
$$+(D_2-\lambda)\sqrt{mn}f(m)f(n)\rho_{m-1,n-1}$$
$$-D_1\sqrt{m(n+1)}f(m)f(n+1)\rho_{m-1,n+1}$$
$$-D_1^*\sqrt{m+1}\,n\,f(m+1)f(n)\rho_{m+1,n-1}$$
$$+\frac{1}{2}D_1\sqrt{(n+1)(n+2)}f(n+1)f(n+2)\rho_{m,n+2}$$
$$+\frac{1}{2}D_1\sqrt{m(m-1)}f(m-1)f(m)\rho_{m-2,n}$$
$$+\frac{1}{2}D_1^*\sqrt{(m+1)(m+2)}f(m+1)f(m+2)\rho_{m+2,n}$$
$$+\frac{1}{2}D_1^*\sqrt{n(n-1)}f(n-1)f(n)\rho_{m,n-2}. \quad (46)$$

Here we used the abbreviated notation $\rho_{mn} = <m|\rho(t)|n>$. This equation, complicated in form and in indices involved, comprises not only the density matrix elements in symmetrical forms, such as $\rho_{m\pm 1, n\pm 1}$, but also those matrix elements in asymmetrical forms like $\rho_{m\pm 2,n}, \rho_{m,n\pm 2}$ and $\rho_{m\mp 1,n\pm 1}$. Eq. (46) gives an infinite hierarchy of coupled equations for the off-diagonal matrix elements. Their evolution in time depends on the diffusion and friction coefficients and on the deformation. When $D_1 = 0$, the diagonal elements are coupled only amongst themselves and not coupled to the off-diagonal elements. In this case the diagonal elements (populations) satisfy a simpler equation:

$$\frac{dP(n)}{dt} = -\{(D_2-\lambda)(n+1)f^2(n+1)+(D_2+\lambda)nf^2(n)\}P(n)$$
$$+(D_2+\lambda)(n+1)f^2(n+1)P(n+1)+(D_2-\lambda)nf^2(n)P(n-1), \quad (47)$$

where we have set $P(n) \equiv \rho_{nn}$. For a $q$-deformation, Eq. (47) takes the form

$$\frac{dP(n)}{dt} = -\{(D_2-\lambda)[n+1]+(D_2+\lambda)[n]\}P(n)$$
$$+(D_2+\lambda)[n+1]P(n+1)+(D_2-\lambda)[n]P(n-1). \quad (48)$$

We define the transition probabilities

$$t_+(n) = (D_2 - \lambda)(n+1)f^2(n+1), \quad t_-(n) = (D_2 + \lambda)nf^2(n), \tag{49}$$

which for a $q$-deformation look like

$$t_+(n) = (D_2 - \lambda)[n+1], \quad t_-(n) = (D_2 + \lambda)[n]. \tag{50}$$

With these notations Eq. (47) becomes:

$$\frac{dP(n)}{dt} = t_+(n-1)P(n-1) + t_-(n+1)P(n+1) - [t_+(n) + t_-(n)]P(n). \tag{51}$$

The steady state solution of Eq. (51) is found to be

$$P_{ss}(n) = P(0)\left(\frac{D_2 - \lambda}{D_2 + \lambda}\right)^n. \tag{52}$$

We remark that in the steady state the detailed balance condition holds:

$$t_-(n)P(n) = t_+(n-1)P(n-1). \tag{53}$$

In the particular case of a thermal state, when the diffusion coefficients have the form (17), the stationary solution of Eq. (51) takes the following form:

$$P_{ss}^{th}(n) = Z^{-1} \exp\left[-\frac{\hbar\omega}{2kT}(2n+1)\right], \tag{54}$$

with

$$Z^{-1} = P(0) \exp\frac{\hbar\omega}{2kT}, \tag{55}$$

$Z$ being the partition function for the ordinary harmonic oscillator

$$Z = \sum_{n=0}^{\infty} \exp\left[-\frac{\hbar\omega}{2kT}(2n+1)\right] = \frac{1}{2\sinh\frac{\hbar\omega}{2kT}}. \tag{56}$$

The distribution (54) can also be written

$$P_{ss}^{th}(n) = Z^{-1} \exp\left(-\frac{E_n}{kT}\right), \tag{57}$$

where $E_n = \hbar\omega(n+1/2)$ is the ordinary expression for the energy of the harmonic oscillator. Expressions (54) and (57) have the known form of the Boltzmann distribution. We notice that in the expressions (52) and (54) of the stationary distribution the deformation of the dissipation does not play any role, as expected, due to the fact that in the considered model the deformation is introduced only in the interaction of the system with the environment and the Hamiltonian of the system is kept undeformed.

# 6   Summary

The Lindblad theory of open systems based on quantum dynamical semigroups provides a selfconsistent treatment of damping as a possible extension of quantum mechanics to open systems. Our purpose was to study the dynamics of the quantum harmonic oscillator in interaction with a dissipative environment, with a $f$- and $q$-deformed type of interaction. We have deformed only the operators in the dissipative part of the Lindblad equation, but not the Hamiltonian itself. In this case the diffusion and dissipation coefficients which model the influence of the environment on the oscillator are constant quantities for a given temperature and do not depend on the introduced nonlinearities (number operator). In the limit $f \to 1$, when deformation disappears, the obtained master equation for the damped oscillator with deformed dissipation becomes the usual master equation for the damped oscillator obtained in the framework of the Lindblad theory. The equations of motion for the observables of the considered system are strongly nonlinear and they form a set of coupled differential equations which is not closed. By truncating the set of coupled differential equations for the expectation values of the number operator and squared number operator, we were able to obtain a closed system of coupled differential equations in the case of a small $q$-deformation. The obtained solutions show explicitly the dependence on the small deformation parameter $\tau$. We have also derived the equation for the density matrix in the number representation and in the case of a thermal bath we obtained the stationary solution, which is of the form of the ordinary Boltzmann distribution, as expected, since we did not deform the harmonic oscillator Hamiltonian.

A more complicated model one obtains by introducing a $f$-deformation for the harmonic oscillator operators, both in the Hamiltonian of the system and in the dissipative part of the master equation. Work is in progress in this direction.

The master equation for the damped harmonic oscillator with deformed dissipation is an operator equation and it could be useful to study its consequences by transforming it into more familiar forms, such as the partial differential equations of Fokker-Planck type for the Glauber, antinormal ordering and Wigner quasiprobability distributions or for analogous deformed quasiprobabilities [8] associated with the density operator. It could also be interesting to find the states which minimize the rate of entropy production, which play an important role in the description of the decoherence induced by the environment.

**Acknowledgements**

The authors thank P. Raychev for fruitful discussions. One of us (A. I.) is pleased to express his sincere gratitude for the hospitality at the Institut für Theoretische Physik in Giessen.